\documentstyle[12pt]{article}
\catcode`\@=11

\@addtoreset{equation}{section}

\def\@normalsize{\@setsize\normalsize{15pt}\xiipt\@xiipt
\abovedisplayskip 14pt plus3pt minus3pt%
\belowdisplayskip \abovedisplayskip
\abovedisplayshortskip  \z@ plus3pt%
\belowdisplayshortskip  7pt plus3.5pt minus0pt}

\def\small{\@setsize\small{13.6pt}\xipt\@xipt
\abovedisplayskip 13pt plus3pt minus3pt%
\belowdisplayskip \abovedisplayskip
\abovedisplayshortskip  \z@ plus3pt%
\belowdisplayshortskip  7pt plus3.5pt minus0pt
\def\@listi{\parsep 4.5pt plus 2pt minus 1pt
            \itemsep \parsep
            \topsep 9pt plus 3pt minus 3pt}}

\def\underline#1{\relax\ifmmode\@@underline#1\else
	$\@@underline{\hbox{#1}}$\relax\fi}
\@twosidetrue

\catcode`@=12

\catcode`\@=11

\def\FERMIPUB{}
\def\FERMILABPub#1{\def\FERMIPUB{#1}}
\def\ps@headings{\def\@oddfoot{}\def\@evenfoot{}
\def\@oddhead{\hbox{}\hfill
	\makebox[.5\textwidth]{\raggedright\ignorespaces --\thepage{}--
	\hfill {\rm FERMILAB--Pub--\FERMIPUB}}}
\def\@evenhead{\@oddhead}
\def\subsectionmark##1{\markboth{##1}{}}
}

\ps@headings

\catcode`\@=12

\relax

\def\figcap{\section*{Figure Captions\markboth
	{FIGURECAPTIONS}{FIGURECAPTIONS}}\list
	{Figure \arabic{enumi}:\hfill}{\settowidth\labelwidth{Figure 999:}
	\leftmargin\labelwidth
	\advance\leftmargin\labelsep\usecounter{enumi}}}
 \relax
\def\tablecap{\section*{Table Captions\markboth
	{TABLECAPTIONS}{TABLECAPTIONS}}\list
	{Table \arabic{enumi}:\hfill}{\settowidth\labelwidth{Table 999:}
	\leftmargin\labelwidth
	\advance\leftmargin\labelsep\usecounter{enumi}}}
 \relax
\def\reflist{\section*{References\markboth
	{REFLIST}{REFLIST}}\list
	{[\arabic{enumi}]\hfill}{\settowidth\labelwidth{[999]}
	\leftmargin\labelwidth
	\advance\leftmargin\labelsep\usecounter{enumi}}}
 \relax

\catcode`@=12
\newskip\humongous \humongous=0pt plus 1000pt minus 1000pt
\def\caja{\mathsurround=0pt}

\newif\ifdtup
\def\panorama{\global\dtuptrue \openup1\jot \caja
	\everycr{\noalign{\ifdtup \global\dtupfalse
	\vskip-\lineskiplimit \vskip\normallineskiplimit
	\else \penalty\interdisplaylinepenalty \fi}}}
\def\eqalignno#1{\panorama \tabskip=\humongous
	\halign to\displaywidth{\hfil$\displaystyle{##}$
	\tabskip=0pt&$\displaystyle{{}##}$\hfil
	\tabskip=\humongous&\llap{$##$}\tabskip=0pt
	\crcr#1\crcr}}

\def\oldreffmt#1{\rlap{[#1]} \hbox to 2\parindent{}}

\def\figfmt#1{\rlap{Figure {#1}} \hbox to 1in{}}

\def\beq{\begin{equation}}
\def\eeq{\end{equation}}

\relax

\begin{document}
\def\tilde{\widetilde}
\def\thefootnote{\fnsymbol{footnote}}
\FERMILABPub{95/107--T}
\begin{titlepage}
\begin{flushright}
        FERMILAB--PUB--95/107--T\\
        OSU Preprint 303\\
        May 1995\\
\end{flushright}
\begin{center}
{\bf An Explicit SO(10) x U(1)$_F$ Model of the Yukawa Interactions}
\vskip 0.20in
        {\bf Carl H. ALBRIGHT}\\
 Department of Physics, Northern Illinois University, DeKalb, Illinois
60115\footnote{Permanent address}\\[-0.2cm]
        and\\[-0.2cm]
 Fermi National Accelerator Laboratory, P.O. Box 500, Batavia, Illinois
60510\footnote{Electronic address: ALBRIGHT@FNALV}\\
\vskip 0.1in
        {\bf Satyanarayan NANDI}\\
 Department of Physics, Oklahoma State University, Stillwater, Oklahoma
        74078\footnote{Electronic address: PHYSSNA@OSUCC}\\
\end{center}
\vfill
\begin{abstract}
\indent We construct an explicit $SO(10) \times U(1)_F$ model of the Yukawa
interactions by using as a guide previous phenomenological results obtained
from a bottom-up approach to quark and lepton mass matrices. The global
$U(1)_F$ family symmetry group sets the textures for the Majorana and generic
Dirac mass matrices by restricting the type and number of Higgs diagrams which
can contribute to each matrix element, while the $SO(10)$ group relates each
particular element of the up, down, neutrino and charged lepton Dirac matrices.
The Yukawa couplings and vacuum expectation values associated with pairs of
{\bf 1,~45, 10,} and {\bf 126} Higgs representations successfully correlate
all the quark and lepton masses and mixings in the scenario incorporating
the nonadiabatic solar neutrino and atmospheric neutrino depletion effects.
\end{abstract}
\end{titlepage}

In a series of manuscripts [1], the authors have demonstrated how a new
bottom-up approach to the quark and lepton mass and mixing problem can be used
to construct phenomenological quark and lepton mass matrices at the
supersymmetric
$SO(10)$ grand unification scale, which lead to the assumed experimental
input at the low scales.   As such, this procedure provides an alternative to
the usual top-down approach [2], where mass matrices are constructed based on
some well-defined theoretical concepts.  Of special interest is a
set of mass matrices found by our approach which exhibit a particularly
simple $SO(10)$ structure for the scenario based on the depletions of solar [3]
and atmospheric [4] neutrinos through oscillations.

In this letter we construct an explicit $SO(10) \times U(1)_F$ model at the
grand unification scale by making use of the phenomenological mass matrices
as a guide.  The global $U(1)_F$ family symmetry singles out a rather
simple set of tree diagrams which set the textures for the Dirac and Majorana
mass matrices, while $SO(10)$ relates the corresponding up, down, neutrino and
charged lepton Dirac matrix elements to each other.  The quantitative
numerical results obtained from the model agree in detail with the input
data assumed for the bottom-up approach.

The starting point for our bottom-up approach was the reasonably well-known
quark mass and Cabbibo-Kobayashi-Maskawa mixing matrix data [5].  To this we
appended neutrino mass and mixing data consistent with the nonadiabatic
Mikheyev-Smirnov-Wolfenstein (MSW) [6]
resonant matter oscillation depletion [3] of the solar electron-neutrino flux
together with atmospheric muon-neutrino depletion [4] through oscillations into
tau-neutrinos.  After running the Yukawa couplings to the grand unification
scale, we applied Sylvester's theorem, as illustrated by Kusenko [7] for quark
data alone, to reconstruct complex-symmetric mass matrices.  The construction
is clearly not unique, but one can vary two parameters which determine the
weak bases in order to select a set of mass matrices which exhibit particularly
simple $SO(10)$ structure for as many matrix elements as possible.  We refer
the interested reader to Ref. [1] for details and begin here with the
special phenomenological matrices singled out by this procedure:
	$$M^U \sim M^{N_{Dirac}} \sim diag(\overline{126};\ \overline{126};
		\ 10)\eqno(1a)$$
	$$M^D \sim M^E \sim \left(\matrix{10',\overline{126} & 10',
		\overline{126}' & 10'\cr 10',\overline{126}' & \overline{126}
		& 10'\cr 10' & 10' & 10\cr}\right)\eqno(1b)$$
with $M^D_{11},\ M^E_{12}$ and $M^E_{21}$ anomalously small and only the
13 and 31 elements complex.  Entries in the matrices stand for the Higgs
representations contributing to those elements.  Recall that the $SO(10)$
product rules read
	$$\eqalignno{
	{\bf 16} \times {\bf 16} &= {\bf 10}_s + {\bf 120}_a + {\bf 126}_s
		&(2a)\cr
        {\bf 16} \times {\bf \overline{16}} &= {\bf 1} + {\bf 45} + {\bf 210}
		&(2b)}$$
We have assumed that vacuum expectation values (VEVs) develop only for the
symmetric
representations ${\bf 10}$ and ${\bf 126}$ and for ${\bf 1}$ and ${\bf 45}$.
The Majorana neutrino mass matrix $M^R$, determined from the seesaw formula [8]
with use of $M^{N_{Dirac}}$ and the reconstructed light neutrino mass matrix,
exhibits a nearly geometrical structure given by [9]
$$M^R \sim \left(\matrix{F & - \sqrt{FE} & \sqrt{FC}\cr
                - \sqrt{FE} & E & -\sqrt{EC}\cr
                \sqrt{FC} & -\sqrt{EC} & C\cr}\right)
                        \eqno(3)$$
where $E = {5\over{6}}\sqrt{FC}$ with all elements relatively real.  It can
not be purely geometrical, however, since the singular rank-1
matrix can not be inverted as required by the seesaw formula,
$M^{N_{eff}} \simeq - M^{N_{Dirac}}(M^R)^{-1}M^{N_{Dirac}^T}$.

The challenge is now to introduce a family symmetry which will enable one
to derive the mass matrix patterns in a simple fashion.  For this purpose,
we propose to use a global $U(1)_F$ family symmetry [10] and to reduce the
problem to the construction of one generic Dirac matrix, $M_{Dirac}$,
along with the single Majorana matrix, ${M^R}$.
As noted above, the $SO(10)$ symmetry will relate the corresponding matrix
elements of the four Dirac matrices to each other.  The important roles
played here by supersymmetry (SUSY) are twofold.  Not only does SUSY control
the running of the Yukawa couplings between the SUSY GUT scale and the
weak scale where it is assumed to be softly broken, but it also allows one to
assume that only simple tree diagrammatic contributions to the mass matrices
need be considered as a result of the nonrenormalization theorem applied
to loop diagrams.  This tree diagram procedure was first suggested by
Dimopoulos [11] twelve years ago.

Simplicity of the $SO(10)$ structure requires that just one Higgs ${\bf 10}$
representation contributes to the $(M_{Dirac})_{33}$ element (hereafter labeled
D33),
i.e., we assume complete unification of the Yukawa couplings at the unification
scale: $\bar{m}_{\tau} = \bar{m}_b = \bar{m}_t/\tan \beta_{10}$, where
$\tan \beta_{10}$ is equal to the ratio of the up quark to the down quark
VEVs in the ${\bf 10}$
$$\begin{array}{rl}
	\bar{m}_t&= g_{10}(v/\surd{2})\sin \beta_{10} \equiv g_{10}v_u
	\nonumber\\
	\bar{m}_b = \bar{m}_{\tau}&= g_{10}(v/\surd{2})\cos \beta_{10}
		\equiv g_{10}v_d \cr
	\tan \beta_{10}&= v_u(5)/v_d(\bar{5})\cr
	\end{array} \eqno(4a)$$
in terms of the $SU(5)$ decomposition of $SO(10)$ with $v = 246$ GeV.
The same ${\bf 10}$ can not contribute to D23 = D32, for the diagonal nature
of $M^U$ and $M^{N_{Dirac}}$ requires the presence of another
${\bf 10'}$ with
	$$ \tan \beta_{10'} = v'_u(5')/v'_d(\bar{5'}) = 0 \eqno(4b)$$
Likewise we assume a pure ${\bf \overline{126}}$ contribution to D22 with
	$$ \tan \beta_{\overline{126}} = w_u(5)/w_d(\overline{45}) \eqno(4c)$$
The tree diagrams for these $M_{Dirac}$ matrix elements are illustrated
in Fig. 1a.

We shall now assign $U(1)_F$ charges to the three families (in order of
appearance) and to the Higgs representations as follows with the numerical
values to be determined later:
	$${\bf 16}_3^{\alpha},\ {\bf 16}_2^{\beta},\ {\bf 16}_1^{\gamma},
		\ {\bf 10}^a, \ {\bf 10'}^b,\ {\bf \overline{126}^c} \eqno(5a)$$
Conservation of $U(1)_F$ charges then requires $2\alpha + a = 0,\ \alpha +
\beta + b = 0$ and $2\beta + c = 0$ as seen from the diagrams in Fig. 1a.

In the above, we have taken the 2-3 sector of $M_{Dirac}$ to be
renormalizable with two ${\bf 10}$'s and one ${\bf \overline{126}}$ developing
low scale VEVs.  We assume the rest of the $M_{Dirac}$ elements arise from
non-renormalizable contributions with the leading ones shown in Fig. 1b.
For D13 we introduce a ${\bf 45}_X^e$ Higgs
field and construct an explicitly complex-symmetric contribution with the
dimension-6 diagram, for which $U(1)_F$ charge conservation
requires $\alpha + \gamma + b + 2e = 0$.  This ${\bf 45}_X$ Higgs field
develops a VEV in
the direction which breaks $SO(10) \rightarrow SU(5) \times U(1)_X$ with
the $SU(5)$ subgroup remaining unbroken.  For D12 we introduce a different
${\bf 45}_Z^h$ Higgs field which breaks $SO(10) \rightarrow {\rm flipped}\
SU(5) \times U(1)$ and is related to the orthogonal ${\bf 45}_X$ and
${\bf 45}_Y$ hypercharge VEVs by
	$$<{\bf 45}_Z> = {6\over{5}}<{\bf 45}_X> - {1\over{5}}<{\bf 45}_Y>
		\eqno(6)$$
as given in Table I.  This Higgs field contributes to
$M^D_{12}$ but not to $M^E_{12}$; thus it generates a zero in this position for
the charged lepton mass matrix as suggested in (1b).  The $U(1)_F$ charge
conservation equation
reads $\beta + \gamma - b + 2h = 0$, as the ${\bf 10'}^*$ Higgs field is
required here to reduce the number of contributing diagrams.  The D11 element
is dimension-8 or higher and is left unspecified. The complex-symmetric Yukawa
diagrams which we wish to generate are then neatly summarized by the ordering
of the Higgs fields:
	$$\begin{array}{rl}
	  D33:&  {\bf 16}_3 \ {\bf - 10 - 16}_3\nonumber \\
	  D23:&  {\bf 16}_2 \ {\bf - 10' - 16}_3\\
	  D32:&  {\bf 16}_3 \ {\bf - 10' - 16}_2\\
	  D22:&  {\bf 16}_2 \ {\bf - \overline{126} - 16}_2\\
	  D13:&  {\bf 16}_1 \ {\bf - 45}_X\ {\bf- 10' - 45}_X\ {\bf - 16}_3\\
	  D31:&  {\bf 16}_3 \ {\bf - 45}_X\ {\bf - 10' - 45}_X\ {\bf - 16}_1\\
	  D12:&  {\bf 16}_1 \ {\bf - 45}_Z\ {\bf - 10'^* - 45}_Z\ {\bf - 16}_2\\
	  D21:&  {\bf 16}_2 \ {\bf - 45}_Z\ {\bf - 10'^* - 45}_Z\ {\bf - 16}_1\\
	  \end{array} \eqno(7a)$$

In order to obtain a different set of diagrams for the Majorana
matrix, we begin the M33 contribution with a dimension-6 diagram shown in
Fig. 1c by including a new ${\bf \overline{126}'}^d$ Higgs which develops a VEV
at the GUT scale in the $SU(5)$ singlet direction, along with a pair of
${\bf 1}^g$ Higgs fields.  Here $2\alpha + d + 2g = 0$.  The nearly geometric
structure for $M^R$ can then be generated by appending more
Higgs fields to each diagram.  For M23 we introduce another ${\bf 1'}^f$ Higgs
field to construct a diagram with one ${\bf \overline{126}'}^d$, one
${\bf 45}_X^e$, one ${\bf 1'}^f$ and two ${\bf 1}^g$ fields with charge
conservation demanding $\alpha + \beta + d + 2g + e + f = 0$.  The new
${\bf 1'}$ field is needed in order to scale properly the Majorana matrix
elements relative to each other.  The remaining leading-order diagrams of the
complex-symmetric Majorana mass matrix
follow by appending more ${\bf 45}_X^e$, ${\bf 45}_Z^h$ and ${\bf 1'}^f$
Higgs lines.  The pattern is made clear from the charge conservation equations:
$2\beta + d + 2g + 2e + 2f =0$ for M22, $\alpha + \gamma + d + 2g + e + h
+ 2f = 0$ for M13, $\beta + \gamma + d + 2g + 2e + h + 3f = 0$ for M12,
and $2\gamma + d + 2g + 2e + 2h + 4f = 0$ for M11.

In summary, the following
Higgs representations have been introduced in addition to those in (5a):
	$${\bf \overline{126}}'^d,\ {\bf 45}_X^e,\ {\bf 45}_Z^h,\ {\bf 1}^g,
		\ {\bf 1'}^f \eqno(5b)$$
all of which generate massive VEVs near the GUT scale.  In order to obtain
CP-violation in the quark and lepton mixing matrices, we allow the VEVs for
${\bf 45}_X,\ {\bf 45}_Z,\ {\bf 1}$ and ${\bf 1'}$ to be complex, but the
VEVs associated with the ${\bf 10,\ 10'\ , \overline{126}}$ and ${\bf
\overline{126}'}$ representations can be taken to be real without loss of
generality as seen from our bottom-up results.
Clearly, many permutations of the Higgs fields are possible in the higher-order
diagrams.

At this point a computer search was carried out to generate $U(1)_F$ charge
assignments leading to the fewest additional diagrams allowed by charge
conservation.  An especially interesting charge assignment stood out for which
	$$\begin{array}{rl}
	       &\alpha = 9,\ \beta = -1,\ \gamma = -8 \nonumber \\
	&a = -18,\ b = -8,\ c = 2,\ d = -22,\ e = 3.5,\ f = 6.5,\ g = 2.0,\
	  	h = 0.5\cr
	\end{array} \eqno(8a)$$
One should note that since $\alpha + \beta + \gamma = 0$, the $\left[SO(10)
\right]^2 \times U(1)_F$ anomaly vanishes, whereas the $\left[U(1)_F\right]^3$
anomaly does not.  Simplicity then suggests that the $U(1)_F$ family
symmetry group be global with a familon being generated upon its breaking.

With the above charge assignments we can greatly limit the number of
permutations and eliminate other unwanted diagrams by restricting the $U(1)_F$
charges appearing on the superheavy internal fermion lines.  With the following
minimum set of allowed charges for the left-handed superheavy fermions $F_L$
and their mirror partners $F^c_L$
	$$\begin{array}{rrrrrrrrrr}
	F_L:& -0.5,& 1.0,& 2.0,& 4.0,& 4.5,& -4.5,& -7.5,& 11.0,& 12.5 \nonumber
		\\
      F^c_L:& 0.5,& -1.0,& -2.0,& -4.0,& -4.5,& 4.5,& 7.5,& -11.0,& -12.5\cr
	\end{array} \eqno(8b)$$
we recover just the leading-order diagrams listed in (7a) for the generic Dirac
mass matrix together with the following uniquely-ordered diagrams for the
complex-symmetric Majorana mass matrix
	$$\begin{array}{rl}
	  M33:&  {\bf 16}_3 \ {\bf - 1 - \overline{126}' - 1 - 16}_3\nonumber \\
	  M23:&  {\bf 16}_2 \ {\bf - 1 - 45}_X\ {\bf - 1' - \overline{126}' - 1
		 - 16}_3\\
          M32:&  {\bf 16}_3 \ {\bf - 1 - \overline{126}' - 1' - 45}_X\ {\bf - 1
		 - 16}_2\\
	  M22:&  {\bf 16}_2 \ {\bf - 1 - 45}_X\ {\bf - 1' - \overline{126}' -
		1' - 45}_X\ {\bf - 1 - 16}_2\\
	  M13:&  {\bf 16}_1 \ {\bf - 45}_X\ {\bf - 1' - 1- 45}_Z\ {\bf - 1' -
		\overline{126}' - 1 - 16}_3\\
          M31:&  {\bf 16}_3 \ {\bf - 1 - \overline{126}' - 1' - 45}_Z\ {\bf -
		1 - 1' - 45}_X\ {\bf - 16}_1\\
	  M12:&  {\bf 16}_1 \ {\bf - 45}_X\ {\bf - 1' - 1- 45}_Z\ {\bf - 1' -
		\overline{126}' - 1' - 45}_X\ {\bf - 1 - 16}_2\\
          M21:&  {\bf 16}_2 \ {\bf - 1 - 45}_X\ {\bf - 1' - \overline{126}' -
		1' - 45}_Z\ {\bf - 1 - 1' - 45}_X\ {\bf - 16}_1\\
	  M11:&  {\bf 16}_1 \ {\bf - 45}_X\ {\bf - 1' - 1- 45}_Z\ {\bf - 1' -
		\overline{126}' - 1' - 45}_Z\ {\bf - 1 - 1' - 45}_X
		{\bf - 16}_1\\
	  \end{array} \eqno(7b)$$
Several other higher-order diagrams are allowed by the $U(1)_F$ charges given
in (8a,b) and appear for D11, D22, M23 and M32 with the Higgs fields ordered
as follows:
	$$\begin{array}{rl}
	  D11:&  {\bf 16}_1 \ {\bf - 45}_X\ {\bf - 1' - 1 - \overline{126} - 1
		- 1' - 45}_X\ {\bf - 16}_1\nonumber \\
	  D22:&  {\bf 16}_2 \ {\bf - 45}_Z\ {\bf - 10'^* - 1'^* - 16}_2,\quad
	      	{\bf 16}_2 \ {\bf - 1'^* - 10'^* - 45}_Z\ {\bf - 16}_2\\
	  M23:&  {\bf 16}_2 \ {\bf - 45}_X^*\ {\bf - 1' - 1 - 45}_Z\ {\bf - 1'
		- \overline{126}' - 1 - 16}_3\\
          M32:&  {\bf 16}_3 \ {\bf - 1 - \overline{126}' - 1' - 45}_Z\ {\bf -
		1 - 1' - 45}_X^*\ {\bf - 16}_2\\
	  \end{array} \eqno(7c)$$
These corrections to M23 and M32 ensure that $M^R$ is rank 3 and nonsingular,
so that the seesaw formula can be applied.  Up to this point the contributions
are all complex-symmetric.

Additional correction terms of higher order which need not be complex-symmetric
can be generated for the Dirac and Majorana matrix elements, if one allows
additional superheavy fermion pairs with new $U(1)_F$ charges.  Such a subset
which does not destroy the pattern constructed above but helps to improve the
numerical results given later consists of the following:
	$$\begin{array}{rrrrrr}
	F_L:&  1.5,& -6.0,& -6.5 \nonumber \\
      F^c_L:& -1.5,&  6.0,&  6.5 \cr
	\end{array} \eqno(8c)$$
We shall enumerate the additional diagrams contributing to D11, D12, D13,
D21, D31 and M11 in a more detailed paper in preparation [12].

We assume the superheavy fermions all get massive at the same mass scale, so
each ${\bf 1, 1', 45}_X$ or ${\bf 45}_Z$ vertex factor can be rescaled by the
same propagator mass $M$.  As a result there are 14 independent parameters
which can then be taken to be
	$$\begin{array}{rl}
	  &g_{10}v_u,\ g_{10}v_d,\ g_{10'}v'_d,\ g_{126}w_u,\ g_{126}w_d,
		\ g_{126'}w' \nonumber\\
	  &g_{45_X}u_{45_X}/M,\ g_{45_Z}u_{45_Z}/M,\ g_{1}u_1/M,
		\ g_{1'}u_{1'}/M \cr
	  &\phi_{45_X},\ \phi_{45_Z},\ \phi_{1},\ \phi_{1'}  \cr
	  \end{array} \eqno(9)$$
In addition, one needs the Clebsch-Gordan coefficient appearing at each vertex
which can be read off from Table I.  The algebraic contributions to each matrix
element of the four Dirac and one Majorana matrices will be spelled out
explicitly in Ref. [12].

One particularly good numerical choice for the parameters at the SUSY GUT scale
is given by\\
	$$\begin{array}{rlrlrlrl}
	 g_{10}v_u &= 120.3, &g_{10}v_d &= 2.46, &g_{10'}v'_d
		&\multicolumn{2}{l}{= 0.078\ {\rm GeV}}\nonumber\\
	 g_{126}w_u &= 0.314, &g_{126}w_d &= - 0.037, &g_{126'}w'
		&\multicolumn{2}{l}{= 0.8 \times 10^{16}\ {\rm GeV}}\cr
	 g_{45_X}u_{45_X}/M &= 0.130, &g_{45_Z}u_{45_Z}/M &= 0.165,
		&g_1u_1/M &= 0.56, &g_{1'}u'_1/M &= - 0.026\cr
	 \phi_{45_X} &= 35^o, &\phi_{45_Z} &\multicolumn{2}{l}{= \phi_{1} =
		\phi_{1'} = - 5^o} \cr
	  \end{array} \eqno(10)$$
which reduces the number of independent parameters to 12
and leads to the following mass matrices at the SUSY GUT scale\\
$$\eqalignno{\noindent
	M^U&= \left(\matrix{ -0.0010 - 0.0001i& 0.0053 + 0.0034i& -0.0013\cr
	0.0053 + 0.0034i& 0.314 & 0 \cr
	-0.0013 & 0 & 120.3 \cr}\right)&(11a)\cr
\noindent M^D&= \left(\matrix{ -0.0001 & -0.0104 + 0.0004i&
	-0.0029 - 0.0045i \cr
	-0.0077 + 0.0018i & -0.036 & 0.078 \cr
	-0.0033 - 0.0048i & 0.078 & 2.460 \cr}\right)&(11b)\cr
\noindent M^N&= \left(\matrix{ 0.0030 + 0.0003i & -0.079 - 0.051i & 0.0038 \cr
	0.048 + 0.031i & -0.942 & 0 \cr
	0.0038 & 0 & 120.3 \cr}\right)&(11c)\cr
\noindent M^E&= \left(\matrix{ 0.0004  & -0.0020 - 0.0010i &
	-0.0023 - 0.0045i \cr
	0.0060 + 0.0031i & 0.108 & 0.078 \cr
	-0.0009 - 0.0037i & 0.078 & 2.460 \cr}\right)&(11d)\cr
\noindent M^R&= \left(\matrix{(-.075 + .637i)\times 10^{9}&
	(-.139 - .117i)\times 10^{11}& (.106 + .019i)\times 10^{13}\cr
	(-.139 - .117i)\times 10^{11}& (.454 + .541i)\times 10^{12}&
	(-.387 - .152i)\times 10^{14}\cr
	(.106 + .019i)\times 10^{13}& (-.387 - .152i)\times 10^{14}&
	(.247 - .044i)\times 10^{16}\cr}\right)&(11e)\cr}$$
in units of GeV.  The Dirac mass matrix elements appear in the form
$\psi_L^T C^{-1}M(\psi^c)_L$, while the Majorana matrix elements refer to
$(\psi^c)_L^T C^{-1}M(\psi^c)_L$ with $\psi_L$ and $(\psi^c)_L$ each a member
of one of the three families of ${\bf 16}$'s.  Identical contributions also
arise from the transposed Dirac matrices and the right-handed Majorana
matrix.  As such, the true Yukawa couplings $G_Y$ are just half the values of
the $g_Y$'s appearing in (9) and (10).

The masses at the GUT scale can then be found by calculating the eigenvalues
of the Hermitian product $MM^{\dagger}$ in each case, while the mixing matrices
$V_{CKM}$ and $V_{lepton}$ can be calculated with the projection operator
technique of Jarlskog [13].  After evolving these quantities to the low scale,
we find in the quark sector
$$\begin{array}{rlrl}
        m_u(1 {\rm GeV})&= 5.0\ (5.1)\ {\rm MeV},& \qquad m_d(1 {\rm GeV})&=
		7.9\ (8.9)\ {\rm MeV}\nonumber\\
        m_c(m_c)&= 1.27\ (1.27)\ {\rm GeV},& \qquad m_s(1 {\rm GeV})&= 169
		\ (175)\ {\rm MeV}\cr
        m_t(m_t)&= 150\ (165)\ {\rm GeV},& \qquad m_b(m_b)&= 4.09\ (4.25)
		\ {\rm GeV}\cr
  \end{array}\eqno(12a)$$
where we have indicated the preferred values in parentheses.  The
mixing matrix is given by
$$V_{CKM} = \left(\matrix{0.972 & 0.235 & 0.0037e^{-i124^o}\cr
                -0.235 & 0.971 & 0.041\cr
                0.012 & -0.039 & 0.999\\[-0.15in]\cr
		\quad -0.003i & \quad -0.001i & \cr}\right) \eqno(12b)$$
Note that $|V_{ub}/V_{cb}| = 0.090$ with the CP-violating phase $\delta =
124^o$, while $m_d/m_u = 1.59,\\
\ m_s/m_d = 21.3$, cf. Ref. [14].  In the lepton
sector we obtain
$$\begin{array}{rlrl}
        m_{\nu_e}&= 0.10\ (?) \times 10^{-4}\ {\rm eV},& \qquad m_e&=
		0.44\ (0.511)\ {\rm MeV}\nonumber\\
        m_{\nu_{\mu}}&= 0.29\ (0.25) \times 10^{-2}\ {\rm eV},& \qquad m_{\mu}
		&= 99\ (105.5)\ {\rm MeV}\cr
        m_{\nu_{\tau}}&= 0.11\ (0.10)\ {\rm eV},& \qquad m_{\tau}&= 1.777
		\ (1.777)\ {\rm GeV}\cr \end{array}\eqno(13a)$$
and
$$V_{lept} = \left(\matrix{0.998 & 0.050 & 0.038e^{-i122^o}\cr
                -0.036  & 0.873 & 0.486\qquad\cr
                0.043   & -0.485 & 0.873\\[-0.15in]\cr
	        \quad -0.037i & \quad -0.002i & \cr}\right) \eqno(13b)$$
The heavy Majorana neutrino masses are
	$$M^R_1 = 0.63 \times 10^9\ {\rm GeV},\qquad M^R_2 = 0.39 \times
		10^{11}\ {\rm GeV},\qquad M^R_3 = 0.25 \times 10^{16}\ {\rm
		GeV} \eqno(13c)$$
The neutrino masses and mixings are in the correct ranges to explain the
nonadiabatic solar neutrino depletion [3] with small mixing and the atmospheric
neutrino depletion [4] with large mixing:
	$$\begin{array}{rlrl}
	  \delta m^2_{12}&= 8.5 \times 10^{-6}\ {\rm eV^2},&\qquad
		\sin^2 2\theta_{12}&= 1.00 \times 10^{-2}\nonumber\\
	  \delta m^2_{23}&= 1.2 \times 10^{-2}\ {\rm eV^2},&\qquad
		\sin^2 2\theta_{23}&= 0.72 \cr \end{array} \eqno(14)$$

For our analysis, the SUSY GUT scale at which the gauge and Yukawa couplings
unify was chosen to be $\Lambda = 1.2 \times 10^{16}$ GeV.  From (4a) and (11)
we find that $g_{10} = 0.69$.  It is interesting to note that if we equate the
$SO(10)$-breaking and lepton number-breaking VEV, $w'$, with $\Lambda$, we
find $g_{126'} = 0.67 \simeq g_{10}$.  Taking into account the remark
following (11e), we note the true Yukawa couplings are $G_{10} \simeq
G_{126'} \simeq 0.33$.  If we further equate $g_1 = g_{10} \simeq g_{126'}$,
and $u_1 = \Lambda$ for the $U(1)_F$-breaking VEV, we find $M = 1.5 \times
10^{16}$ GeV for the masses of the superheavy fermions which condense
with their mirrors.  These values are all very reasonable.

The ${\bf 45}_X$ and ${\bf 45}_Z$ VEVs appear at nearly
the same scale, $2.8 \times 10^{15}$ and $3.5 \times 10^{15}$ respectively, if
one assumes the same Yukawa coupling as above.  On the other hand, if these
VEVs
appear at the unification scale $\Lambda$ the corresponding Yukawa couplings
are smaller than those found above.  In either case, a consequence of their
non-orthogonal breakings is that $SU(5)$ is broken down to $SU(3)_c \times
SU(2)_L \times U(1)_Y$ at the scale in question.  No further breaking is
required until the electroweak scale and the SUSY-breaking scale are reached.

In summary, we have constructed an $SO(10) \times U(1)_F$ model of the
Yukawa interactions with the following features:

\indent (i) The global $U(1)_F$ family symmetry group singles out a rather
simple set of tree diagrams which determines the texture
of the generic Dirac and Majorana mass matrices, while the $SO(10)$ group
relates corresponding matrix elements of the up, down, neutrino and charged
lepton Dirac matrices to each other. \\
\indent (ii) The dominant second and third family Yukawa interactions are
renormalizable and arise through couplings with Higgs in the ${\bf 10,\ 10'}$
and $\bf \overline{126}$ representations of $SO(10)$.  The remaining Yukawa
interactions are of higher order and require couplings of Higgs in the
${\bf \overline{126}}',\ {\bf 1,\ 1',\ 45}_X$ and ${\bf 45}_Z$ representations
which acquire VEVs near the SUSY GUT scale.\\
\indent (iii) The Higgs
which acquire high scale VEVs break the $SO(10) \times U(1)_F$ symmetry down
to the $SU(3)_c \times SU(2)_L \times U(1)_Y$ standard model symmetry.\\
\indent (iv) Although this non-minimal supersymmetric model involves several
Higgs representations,
the runnings of the Yukawa couplings from the GUT scale to the low-energy
SUSY-breaking scale are controlled mainly by the contributions from the
${\bf 10}$, as in the minimal supersymmetric standard model.\\
\indent (v) In terms of 12 input parameters, 15 masses (including the
heavy Majorana masses) and 8 mixing parameters emerge.  The Yukawa couplings
and the Higgs VEVs are numerically feasible and successfully
correlate all the quark and lepton masses and mixings in the scenario which
incorporates the nonadiabatic solar neutrino and atmospheric neutrino
depletion effects.

We shall elaborate further on the numerical details in a paper now in
preparation [12].  Work is also underway to construct a superpotential for the
model presented here.

The authors gratefully acknowledge the Summer Visitor Program and warm
hospitality of the Fermilab Theoretical Physics Department where much of this
research was initiated and carried out.  We thank Joseph Lykken for his
continued interest and comments during the course of this work.  The research
was supported in part by the U. S. Department of Energy.
\newpage
\begin{reflist}
\item	C. H. Albright and S. Nandi, Phys. Rev. Lett. {\bf 73} (1994) 930;
	Report No. Fermilab-PUB-94/061-T and OSU Report No. 286 (to appear in
	Phys. Rev. D); and Report No. Fermilab-PUB-94/119-T and OSU Report No.
	289 (to be published).

\item   H. Georgi and C. Jarlskog, Phys. Lett. {\bf 86B} (1979) 297;
 	J. A. Harvey, P. Ramond and D. B. Reiss, Phys. Lett. {\bf 92B}
	(1980) 309; H. Arason, D. Castan\~{n}o, B. Keszthelyi, S. Mikaelian,
	E. Piard, P. Ramond and B. Wright, Phys. Rev. Lett. {\bf 67} (1991)
	2933; Phys. Rev. D {\bf 46} (1992) 3945;
	S. Dimopoulos, L. J. Hall and S. Raby, Phys. Rev. Lett.
	{\bf 68} (1992) 1984;  Phys. Rev. D {\bf 45} (1992) 4192;
	{\bf 46}, (1992) R4793; {\bf 47} (1993) R3702;  G. F. Giudice,
	Mod. Phys. Lett. A {\bf 7} (1992) 2429; H. Arason, D. Castan\~{n}o,
	P. Ramond and E. Piard, Phys. Rev. D {\bf 47} (1993) 232;
	P. Ramond, R. G. Roberts and G. G. Ross, Nucl. Phys. {\bf B406} (1993)
	19; A. Kusenko and R. Shrock, Phys. Rev. D {\bf 49}, 4962 (1994);
	K. S. Babu and R. N. Mohapatra, Phys. Rev. Lett. {\bf 74} (1995) 2418.

\item	R. Davis et al., Phys. Rev. Lett. {\bf 20} (1968) 1205; in {\it
	Neutrino '88}, ed. J. Schnepp et al. (World Scientific, 1988);
	K. Hirata et al., Phys. Rev. Lett. {\bf 65} (1990) 1297, 1301;
        P. Anselmann et al., Phys. Lett. B {\bf 327} (1994) 377, 390;
	Dzh. N. Abdurashitov et al., Phys. Lett. B {\bf 328} (1994) 234.

\item   K. S. Hirata et al., Phys. Lett. B {\bf 280} (1992) 146; and {\bf 283}
	(1992) 446; R. Becker-Szendy et al., Phys. Rev. Lett. {\bf 69}
	(1992) and Phys. Rev. D {\bf 46} (1992) 3720;
	W. W. M. Allison et al., Report No. ANL-HEP-CP-93-32; Y. Fukuda
	et al., Phys. Lett. B {\bf 335} (1994) 237.

\item   Particle Data Group, M. Aguilar-Benitez et al., Phys. Rev. D {\bf 50}
	(1994) 1173.

\item   S. P. Mikheyev and A. Yu Smirnov, Yad Fiz. {\bf 42} (1985) 1441
                [Sov. J. Nucl. Phys. {\bf 42} (1986) 913]; Zh. Eksp. Teor.
                Fiz. {\bf 91} (1986) 7 [Sov. Phys. JETP {\bf 64}
                (1986) 4]; Nuovo Cimento {\bf 9C} (1986) 17; L. Wolfenstein,
                Phys. Rev. D {\bf 17} (1978) 2369; {\bf 20} (1979) 2634.

\item   A. Kusenko, Phys. Lett. B {\bf 284} (1992) 390.

\item   M. Gell-Mann, P. Ramond, and R. Slansky, in {\it Supersymmetry},
                edited by P. Van Nieuwenhuizen and D. Z. Freedman
                (North-Holland, Amsterdam, 1979); T. Yanagida, Prog. Theor.
                Phys. {\bf B 315} (1978) 66.

\item   The form of $M^R$ presented here differs somewhat from that in Ref.
	[1], for more recent data on the atmospheric depletion effect was
	taken into account.

\item	For recent use of $U(1)_F$ symmetry to generate patterns of fermion
	mass matrices, see L. Ibanez and G. G. Ross, Phys. Lett. B {\bf 332}
	(1994) 100; P. Binetruy and P. Ramond, LPTHE-ORSAY-94-115 preprint;
	H. Dreiner, G. K. Leontaris , S. Lola and G. G. Ross, Nucl. Phys.
	B {\bf 436} (1995) 461.

\item   S. Dimopoulos, Phys. Lett. B {\bf 129} (1983) 417.

\item   C. H. Albright and S. Nandi, (in preparation).

\item   C. Jarlskog, Phys. Rev. D {\bf 35} (1987) 1685; {\bf 36}
        	(1987) 2138; C. Jarlskog and A. Kleppe, Nucl. Phys. {\bf B286}
                (1987) 245.

\item   J. Gasser and H. Leutwyler, Phys. Rep. C {\bf 87} (1982) 77.
\end{reflist}
\newpage
\vspace*{1in}
\begin{center}
\begin{tabular}{c|rrr|c}
	SU(5) & \multicolumn{3}{c|}{VEV Directions} & Flipped SU(5)\\
	Assignments & ${\bf 45}_X$ & ${\bf 45}_Y$ & ${\bf 45}_Z$ &
		Assignments\\ \hline
	$u,\ d$ & 1 & 1 & 1 & $d,\ u$\\
	$u^c$ & 1 & - 4 & 2 & $d^c$\\
	$d^c$ & - 3 & 2 & - 4 & $u^c$\\
	$\nu,\ \ell$ & - 3 & - 3 & - 3 & $\ell,\ \nu$\\
	$\nu^c$ & 5 & 0 & 6 & $e^c$\\
	$e^c$ & 1 & 6 & 0 & $\nu^c$\\ \hline
\end{tabular}

\vspace*{0.25in}
	Table I.  Couplings of the ${\bf 45}$ VEVs to states in the ${\bf 16}$.
\end{center}

\vspace*{1.5in}
\noindent Fig. 1.  Tree-level diagrams for the (a) renormalizable and (b)
leading-order nonrenormalizable contributions to the generic Dirac mass matrix
and for the (c) 33 element of the Majorana mass matrix.
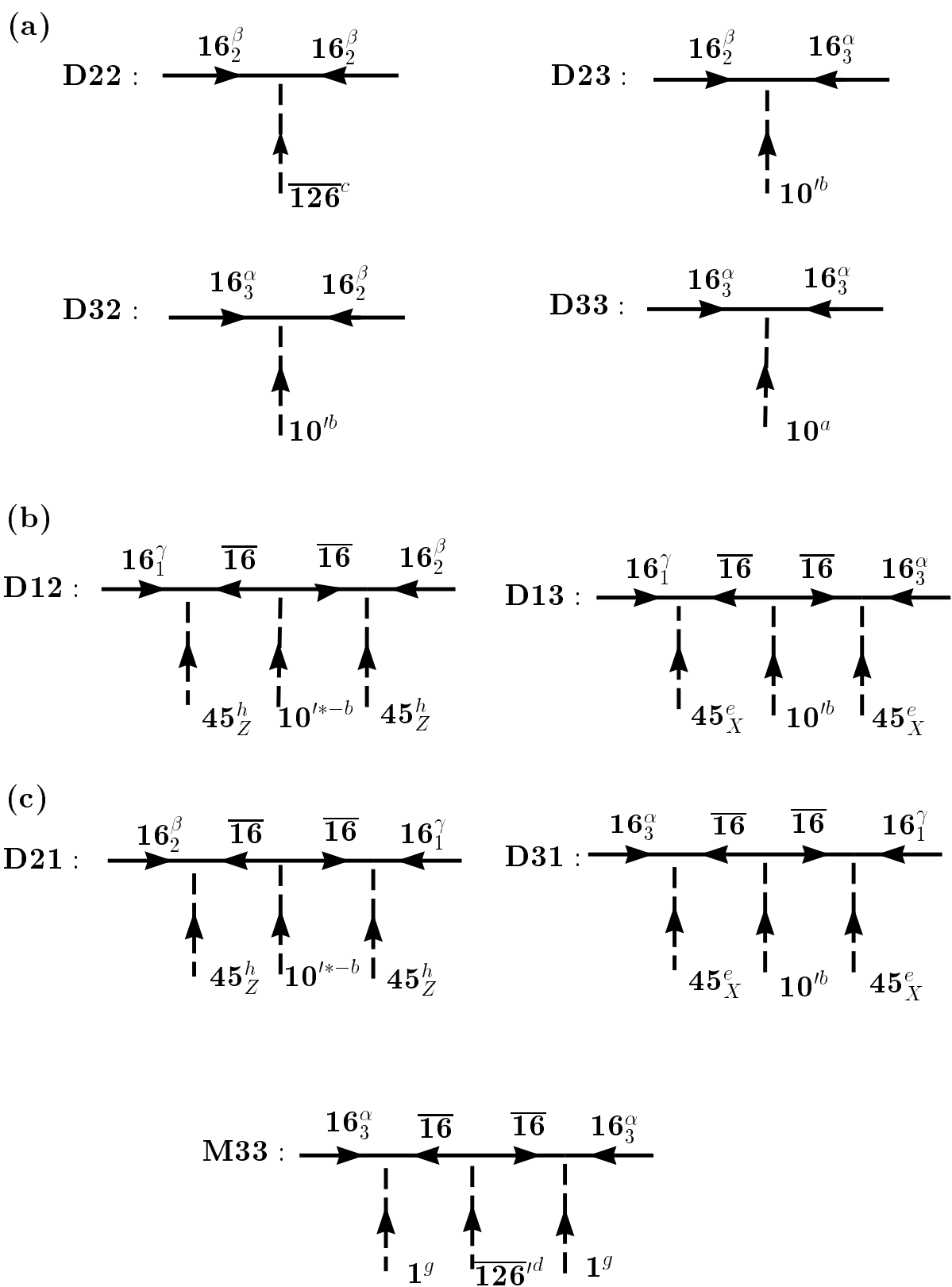
\end{document}